\documentclass[journal=jcisd8,manuscript=article,chaptertitle=true,layout=twocolumn]{achemso}

\usepackage{graphicx}
\usepackage{amsmath}
\usepackage{amssymb}
\usepackage{booktabs}
\usepackage{url}
\usepackage[left]{lineno}
\newcommand{\maybeincludegraphics}[2][]{%
  \IfFileExists{#2}{\includegraphics[#1]{#2}}{%
    \fbox{\parbox[c][0.22\textheight][c]{0.92\linewidth}{\centering Placeholder for figure file \texttt{\detokenize{#2}}}}%
  }%
}

\author{Hodaka Mori}
\email{hodakamori@preferred.jp}
\affiliation[Preferred Networks]{Preferred Networks, Inc., Tokyo 100-0004, Japan}

\author{Yu Miyazaki}
\affiliation[Preferred Networks]{Preferred Networks, Inc., Tokyo 100-0004, Japan}

\author{Takechika Kikkawa}
\affiliation[Preferred Networks]{Preferred Networks, Inc., Tokyo 100-0004, Japan}

\title[CoTAR]{CoTAR: Topology and Atomic State Reconstruction in Condensed Phases}

\begin{document}
\begin{tocentry}
  \centering
  \includegraphics[
    width=\linewidth,height=4.5cm,
    keepaspectratio,clip,
  ]{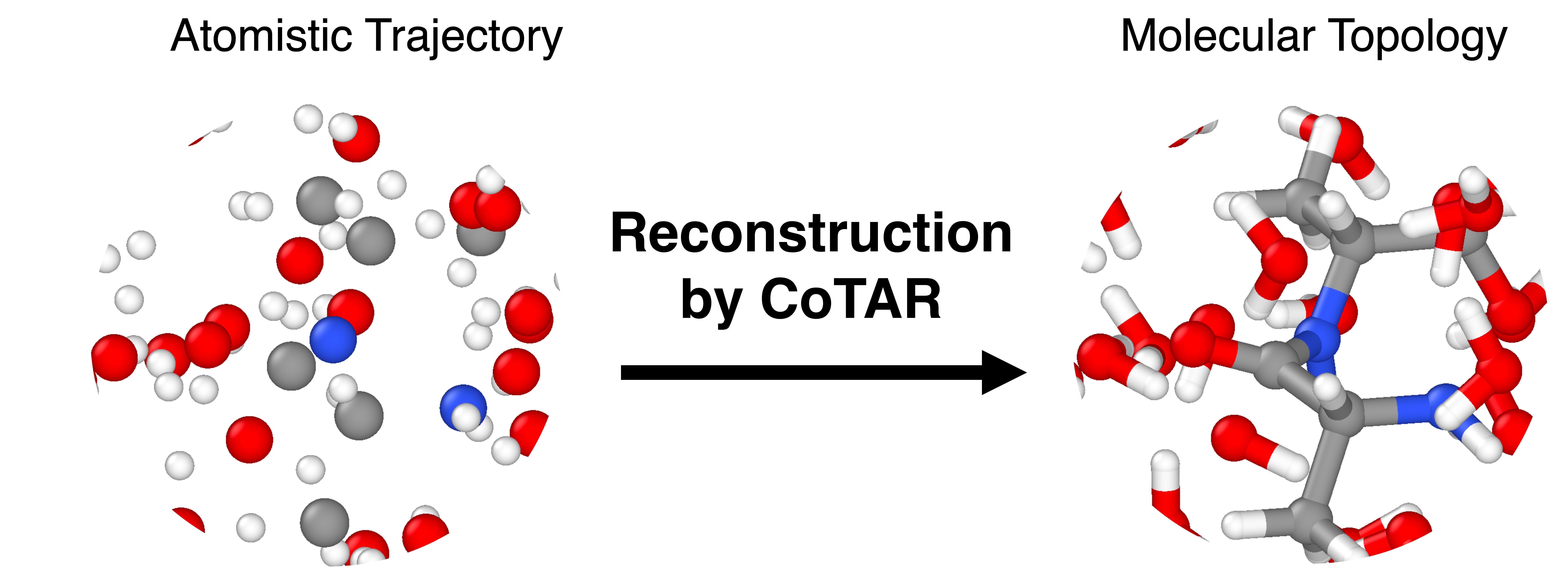}
\end{tocentry}


\begin{abstract}
Universal machine learning interatomic potentials (uMLIPs) enable condensed-phase molecular dynamics (MD) simulations with near-first-principles accuracy, but their lack of explicit molecular topology limits bond-aware analysis and reconnection to classical force fields. Here, we present CoTAR, a hybrid graph neural network (GNN)--hidden Markov model (HMM) framework that reconstructs molecular topology, formal charges, and unpaired electrons from atomic species, coordinates, and total charge by combining message passing on a proximity graph with a van der Waals prior, chemical constraints, and temporal smoothing. Across 128 nonreactive, topology-preserving condensed-phase systems, CoTAR achieved a bond-order-weighted F1 score of 0.906 on classical-MD data; for uMLIP trajectories, few-shot fine-tuning improved the valid-snapshot rate from 38.6\% to 84.7\%. The reconstructed topologies also supported downstream classical MD simulations, and HMM smoothing improved system-level MD simulation feasibility from 83.6\% to 85.9\%, indicating that CoTAR provides a practical framework for bond-aware analysis of condensed-phase uMLIP trajectories.
\end{abstract}

\section{Introduction}

Universal machine learning interatomic potentials (uMLIPs) have rendered condensed-phase molecular dynamics (MD) simulations with near-first-principles accuracy and practical computational cost feasible \cite{Batatia2022mace,Batatia2022Design,Takamoto2022-mj,Chen2022-en,Deng2023-ms}. This capability is important for liquids, solutions, polymers, and interfaces, where quantum chemical treatments are often computationally expensive and classical force fields may lack transferability\cite{wood2025family,simm2025simpoly}.

In contrast to classical force fields, uMLIP-based simulations generally do not require explicit molecular topology\cite{Batatia2022mace,Batatia2022Design,Takamoto2022-mj,Chen2022-en,Deng2023-ms}. This flexibility is advantageous; however, it complicates analyses that depend on bonding information, including molecule identification, reaction tracking, and functional-group analysis. It also complicates the reconnection of selected uMLIP configurations or pathways to classical force fields for larger-scale or longer-timescale simulations.

Distance heuristics are widely used to infer covalent bonds from coordinates\cite{humphrey1996vmd,stukowski2010visualization,zhang2012rule}; however, condensed phases contain many short intermolecular contacts, including hydrogen bonds, solvation-shell contacts, ion pairs, and dense packing contacts. Consequently, these methods can generate false intermolecular bonds and do not assign bond orders. Constraint-based bond-order assignment methods can recover bond orders by enforcing valence and charge consistency on a candidate connectivity graph\cite{kim2015universal}. However, in dense periodic systems, numerous short intermolecular contacts render the candidate graph larger and more ambiguous, which can increase the search cost and lead to practical failures.

Graph neural networks (GNNs) and temporal smoothing schemes such as hidden Markov models (HMMs) have been applied to bond inference, but most reported approaches focus on isolated molecules or dilute systems\cite{loschen,zeng2020reacnetgenerator,magedov2021bond,wang2026multimodal}. Condensed-phase trajectories remain challenging because close intermolecular contacts must be distinguished from covalent bonds under periodic boundary conditions (PBCs) while preserving the consistency of formal charges and unpaired electrons.

Here, we present CoTAR (Coordinates-to-Topology and Atom-State Reconstruction), a hybrid GNN--HMM framework for reconstructing molecular 
topology and atom states from condensed-phase trajectories under PBCs.
CoTAR predicts pair states and atom states from atomic species, coordinates, and total charge, and combines a van der Waals (vdW) prior with valence and charge
constraints to suppress spurious intermolecular bonds and chemically inconsistent assignments. For uMLIP trajectories, we use same-system few-shot 
fine-tuning to address the distribution shift from classical-MD trajectories to uMLIP trajectories. This strategy provides a practical adaptation route 
when a limited number of target-system snapshots can be annotated, without requiring exhaustive reannotation of the full trajectory. We then evaluate 
downstream utility by reconnecting the reconstructed topologies to classical MD.
By integrating learned local geometric context, explicit chemical constraints, atom-state prediction,
and temporal smoothing, CoTAR provides a practical framework for converting topology-free condensed-phase trajectories into chemically complete representations for bond-aware analysis and force-field reconnection. 
The present study focuses on nonreactive, topology-preserving systems, which define the current domain of applicability.
\section{Methods}

\subsection{Overview of the CoTAR framework}

The goal is to infer bond topology and atom states from atomic species $Z_i$, 3D coordinates $\mathbf{r}_i$, and, optionally, the total charge $Q_{\mathrm{total}}$ while satisfying valence and charge constraints.
In condensed phases, purely distance-based decisions are prone to false positives because of dense packing and thermal fluctuations.
Bond order, formal charge, and unpaired-electron count are coupled through electron-count and valence rules; hence, separate predictions readily produce chemically inconsistent assignments.
CoTAR therefore combines (i) message-passing neural network (MPNN)-based pair-state prediction on a proximity graph, (ii) joint prediction of formal charges $q_i$ and unpaired electrons $u_i$, (iii) a vdW prior and chemical constraints, and (iv) HMM-based temporal refinement for trajectories\cite{eddy1996hidden,gilmer2020message}.
\begin{figure}[t]
  \centering
  \maybeincludegraphics[width=\columnwidth]{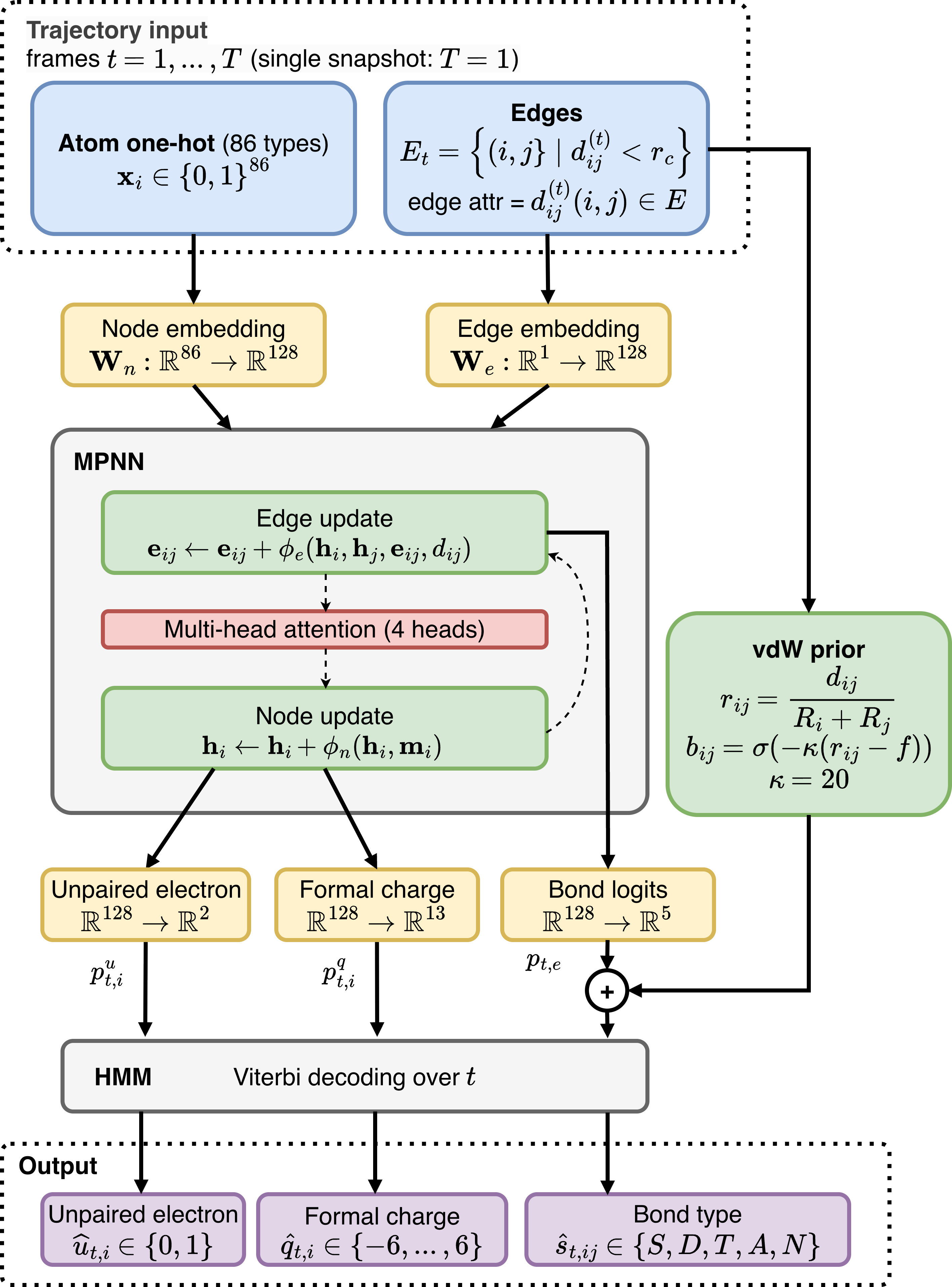}
  \caption{Overview of CoTAR. A proximity graph is constructed for each frame; an MPNN jointly predicts pair states and atom states; a vdW prior biases covalent versus noncovalent assignments; and HMM smoothing enforces temporal consistency for trajectory inputs. During training, additional chemical-constraint losses encourage chemically consistent predictions.}
  \label{fig:cotar_model}
\end{figure}

The input graph $\mathcal{G}=(\mathcal{V},\mathcal{E})$ is constructed by treating atoms as nodes and defining proximity pairs within a distance cutoff of $r_c=3.0~\text{\AA}$.
To distinguish the chemical bond itself from the proximity relation handled by the GNN, we first define the undirected set of proximity pairs $\mathcal{P}$ and then introduce a directed edge set $\mathcal{E}$ for computation:
\begin{align}
\mathcal{V} &= \{1,\dots,N\}, \label{eq:V}\\
\mathcal{P} &= \left\{\{i,j\}\,\middle|\,1\le i<j\le N,\ d_{ij}<r_c\right\}, \label{eq:P}\\
\mathcal{E} &= \left\{(i,j)\,\middle|\, i\neq j,\ \{i,j\}\in\mathcal{P}\right\}, \label{eq:E}\\
d_{ij} &= \|\mathbf{r}_i-\mathbf{r}_j\|. \label{eq:dij}
\end{align}
Under PBCs, $d_{ij}$ is defined by the minimum-image distance.
We denote the proximity neighborhood of atom $i$ by $\mathcal{N}(i)=\{j\mid (i,j)\in\mathcal{E}\}$.
Pair labels are defined for undirected atom pairs and copied to both directed edges.
At inference time, the outputs for $(i,j)$ and $(j,i)$ are averaged to recover a symmetric undirected prediction.

Node features comprise only the element one-hot vector $\mathbf{x}_i\in\{0,1\}^{86}$, where the 86 channels correspond to element types with atomic numbers $Z=1,\dots,86$ (H through Rn). Edge features comprise only the interatomic distance $d_{ij}\in\mathbb{R}$. This minimal representation preserves translational and rotational invariance and applies uniformly to molecules, condensed phases, and periodic systems.

The state space for proximity pairs is defined as
\begin{align}
\mathcal{C}_{\mathrm{cov}} &= \{\mathrm{Single},\mathrm{Double},\mathrm{Triple},\mathrm{Aromatic}\},\\
\mathcal{C} &= \{\varnothing\}\cup\mathcal{C}_{\mathrm{cov}},\qquad |\mathcal{C}|=5,
\end{align}
where $\varnothing$ denotes a noncovalent state in which a pair is spatially close but does not form a covalent bond.
Treating Aromatic as an explicit class absorbs the nonuniqueness of Kekul\'{e} representations in aromatic rings.
In what follows, we refer to these as \emph{pair states}.

\subsection{GNN-based pair-state prediction}

Covalent bonding is not determined by pair distance alone but depends on many-body context such as coordination number, local density, neighboring element composition, and aromaticity.
CoTAR therefore uses message passing to repeatedly aggregate local environmental information and infer pair states conditioned on the surrounding structure.
Condensed phases often contain many neighbors within the cutoff;  hence, multistep propagation is important for distinguishing close contacts from true covalent bonds.
We use $L=16$ message-passing layers to ensure a sufficiently large receptive field. Residual connections together with LayerNorm stabilize deep training.
Neighbor contributions are weighted by 4-head attention.

With hidden dimension $D=128$, the initial node and edge embeddings are defined by linear maps:
\begin{align}
\mathbf{h}_i^{(0)} &= \mathbf{W}_n \mathbf{x}_i, \label{eq:init_node}\\
\mathbf{e}_{ij}^{(0)} &= \mathbf{W}_e d_{ij}, \label{eq:init_edge}\\
\mathbf{W}_n &: \mathbb{R}^{86} \to \mathbb{R}^{D}, \label{eq:map_node}\\
\mathbf{W}_e &: \mathbb{R} \to \mathbb{R}^{D}. \label{eq:map_edge}
\end{align}

At each layer $l=0,\ldots,L-1$, the edge feature is first updated as
\begin{equation}
\mathbf{e}_{ij}^{(l+1)}=
\mathbf{e}_{ij}^{(l)}+
\phi_e\!\left([\mathbf{h}_i^{(l)},\mathbf{h}_j^{(l)},\mathbf{e}_{ij}^{(l)},d_{ij}]\right).
\end{equation}
The distance $d_{ij}$ is provided explicitly at every layer to preserve geometric scale information through the depth of the network.
Messages are then aggregated with 4-head attention:
\begin{align}
\alpha_{ij}^{(l,h)}
&=
\mathrm{softmax}_{j\in\mathcal{N}(i)}
\phi_{\mathrm{att}}^{(h)}\!\left(\mathbf{e}_{ij}^{(l+1)}\right), \label{eq:att}\\
\mathbf{m}_i^{(l)}
&=
\mathrm{Concat}_{h=1}^{4}
\sum_{j\in\mathcal{N}(i)}
\alpha_{ij}^{(l,h)}
\mathbf{W}_m^{(h)}
\mathbf{e}_{ij}^{(l+1)}. \label{eq:message}
\end{align}
Finally, the node feature is updated as
\begin{equation}
\mathbf{h}_i^{(l+1)}=
\mathbf{h}_i^{(l)}+
\phi_n\!\left([\mathbf{h}_i^{(l)},\mathbf{m}_i^{(l)}]\right).
\end{equation}
Both $\phi_e$ and $\phi_n$ are 2-layer MLPs with LayerNorm and SiLU.

The final edge representation is used to predict pair-state logits:
\begin{align}
\mathbf{z}_{ij} &= \mathbf{W}_{\mathrm{pair}}\mathbf{e}_{ij}^{(L)}+\mathbf{b}_{\mathrm{pair}},\\
\hat{\mathbf{p}}_{ij} &= \mathrm{softmax}(\tilde{\mathbf{z}}_{ij}),\\
\hat p_{ij}(c) &= (\hat{\mathbf{p}}_{ij})_c,\qquad c\in\mathcal{C}.
\end{align}
Here $\hat p_{ij}(c)$ is the predicted probability that the proximity pair $\{i,j\}$ belongs to state $c$.

At the same time, the final node representation predicts formal charges $q_i\in\{-6,\ldots,6\}$ and unpaired electrons $u_i\in\{0,1\}$:
\begin{equation}
\mathbf{z}_i^{q}=\phi_q(\mathbf{h}_i^{(L)}),\qquad
\mathbf{z}_i^{u}=\phi_u(\mathbf{h}_i^{(L)}).
\end{equation}
As pair states and $(q_i,u_i)$ are coupled through valence constraints, joint atom-state prediction acts as a chemical regularizer that suppresses false bonds accompanied by valence violations.

A physically motivated vdW prior is further incorporated into the logits\cite{mendeleev2014}.
Here, $R_i$ denotes the element-specific van der Waals radius of atom $i$. The values used in this work were taken from the Mendeleev database\cite{mendeleev2014}.
Defining the distance ratio $r_{ij}=d_{ij}/(R_i+R_j)$, we map it into $b_{ij}\in(0,1)$ by a sigmoid:
\begin{equation}
b_{ij}=\sigma\!\left(-\kappa(r_{ij}-f)\right),\qquad \kappa=20.
\end{equation}
The threshold-like parameter $f$ (initialized to 0.6) and scale $\gamma$ (initialized to 100) are learnable.
This prior acts primarily on the separation between covalent states and the noncovalent state $\varnothing$, while bond-order discrimination is left to the MPNN representation:
\begin{align}
\tilde{z}_{ij}(c)
&=
z_{ij}(c)+\gamma\,b_{ij},
\qquad c\in\mathcal{C}_{\mathrm{cov}},\\
\tilde{z}_{ij}(\varnothing)
&=
z_{ij}(\varnothing)+\gamma(1-b_{ij}).
\end{align}
\subsection{HMM-based temporal refinement}

In MD simulations, coordinates fluctuate continuously, whereas pair states usually do not change abruptly between adjacent frames.
Frame-independent predictions therefore tend to contain high-frequency noise, including spike-like state flips in borderline situations.
To suppress such noise while preserving meaningful state changes, we apply an HMM to the sequence of CoTAR output probabilities and infer the most likely hidden-state trajectory by the Viterbi algorithm\cite{seshadri1994list}.
No smoothing is applied for a single frame ($T=1$).

For time $t$ and proximity pair $e$, let the observation be $\hat{\mathbf{p}}_{t,e}\in[0,1]^{|\mathcal{C}|}$ and the hidden state be $z_{t,e}\in\{1,\ldots,|\mathcal{C}|\}$.
Transitions are controlled by a self-transition probability $p_{\mathrm{self}}$:
\begin{align}
A_{kk}&=p_{\mathrm{self}},\\
A_{k\ell}&=\frac{1-p_{\mathrm{self}}}{|\mathcal{C}|-1}\qquad (\ell\neq k),\\
\pi_k&=\frac{1}{|\mathcal{C}|}.
\end{align}
The emission log-probability is taken directly from the CoTAR probability with numerical stabilization:
\begin{equation}
\log P(o_{t,e}\mid z_{t,e}=k)=\log\!\left(\hat p_{t,e}(k)+\varepsilon\right),\qquad \varepsilon=10^{-10}.
\end{equation}
The most likely sequence is
\begin{equation}
\hat{z}_{1:T,e}=
\arg\max_{z_{1:T,e}}
\left[
\pi_{z_{1,e}}
\prod_{t=1}^{T}P(o_{t,e}\mid z_{t,e})
\prod_{t=2}^{T}A_{z_{t-1,e},z_{t,e}}
\right].
\end{equation}

To reduce computational cost, we fix the set of pairs to be smoothed as the union of proximity pairs over representative frames rather than over the whole trajectory.
If $N_{\mathrm{hmm}}$ representative frames are selected, we define
\begin{equation}
N_{\mathrm{rep}}=\min(T,N_{\mathrm{hmm}}).
\end{equation}
The representative frames are chosen uniformly over the trajectory.
If a pair $e\in\mathcal{P}_{\mathrm{hmm}}$ no longer satisfies the proximity condition at time $t$, its observation is imputed as a one-hot vector for the noncovalent state $\varnothing$.
The same HMM procedure is applied to formal charges (13 states) and unpaired electrons (2 states).
In this study, we use $p_{\mathrm{self}}=0.99$ and $N_{\mathrm{hmm}}=20$.

\subsection{Loss functions and chemical constraints}

Pair-state classification is local and, by itself, does not guarantee global chemical constraints such as total-charge conservation and valence consistency.
We therefore add atom-level terms that penalize violations of electron-count consistency.
The total loss is
\begin{equation}
\mathcal{L}=\mathcal{L}_{\mathrm{pair}}+\mathcal{L}_{\mathrm{atom}}.
\end{equation}

For each annotated snapshot $(c,t)$, we define the valid directed-edge set and valid atom set as
\begin{align}
\mathcal{E}_{\mathrm{valid}}^{(c,t)}
&=
\left\{(i,j)\,\middle|\, \{i,j\}\in\mathcal{P}_c^{(t)}\right\},\\
V_{\mathrm{valid}}^{(c,t)}
&=
\mathcal{V}_c.
\end{align}
In implementation, the loss is averaged over the union of these sets in a minibatch.

Thus, the pair-state classification loss is the average cross-entropy over the valid directed edges:
\begin{equation}
\mathcal{L}_{\mathrm{pair}}
=
\frac{1}{|\mathcal{E}_{\mathrm{valid}}|}
\sum_{(i,j)\in \mathcal{E}_{\mathrm{valid}}}
\mathrm{CE}(\ell_{ij}^{*},\hat{\mathbf{p}}_{ij}).
\end{equation}
Here $\ell_{ij}^{*}\in\mathcal{C}$ is the reference state assigned to the corresponding undirected proximity pair and copied to both directed edges.
We use standard cross-entropy rather than focal loss or explicit class weighting because the candidate restriction imposed by the cutoff, the physical bias from the vdW prior, and the atom-level terms already mitigate the imbalance effectively.

The atom-level loss is defined as a linear combination of multiple constraints, regularizers, and auxiliary supervision terms:
\begin{equation}
\mathcal{L}_{\mathrm{atom}}=\sum_{\tau\in\mathcal{T}}\lambda_{\tau}\mathcal{L}_{\tau},
\end{equation}
where $\mathcal{T}=\{\mathrm{cons},Q,u,|q|,q\mathrm{CE},u\mathrm{CE}\}$.
Constraint terms are evaluated in expectation to avoid nondifferentiability:
\begin{equation}
\hat{q}_i=\sum_{c=-6}^{6}c\,P(q_i=c),\qquad
\hat{u}_i=\sum_{u\in\{0,1\}}u\,P(u_i=u).
\end{equation}
The formal-charge label space is fixed to $c\in\{-6,\dots,6\}$ to cover the range of charge states that may arise for the element types considered in this work ($Z=1,\dots,86$).

The predicted valence induced by pair states is defined as the sum of expected bond orders:
\begin{align}
v_i^{\mathrm{pred}}
&=
\sum_{j\in\mathcal{N}(i)}\bar{b}_{ij},\\
\bar{b}_{ij}
&=
1\cdot \hat p_{ij}(\mathrm{Single})
+2\cdot \hat p_{ij}(\mathrm{Double})
+3\cdot \hat p_{ij}(\mathrm{Triple})
+1.5\cdot \hat p_{ij}(\mathrm{Aromatic}).
\end{align}
The noncovalent state $\varnothing$ does not contribute to $\bar{b}_{ij}$.
Valence consistency is then enforced by
\begin{align}
v_i^{\mathrm{eff}}
&=
v_i^{\mathrm{std}}-\hat{q}_i-\hat{u}_i,\\
\mathcal{L}_{\mathrm{cons}}
&=
\frac{1}{|V_{\mathrm{valid}}|}
\sum_{i\in V_{\mathrm{valid}}}
\left(v_i^{\mathrm{pred}}-v_i^{\mathrm{eff}}\right)^2.
\end{align}
Here, $v_i^{\mathrm{std}}$ denotes the default standard valence associated with element $Z_i$.
Representative values for the principal main-group elements appearing in this study are H=1, C=4, N=3, O=2, F/Cl/Br/I=1, P=3, and S=2.

Total-charge consistency is imposed as a batch-level term:
\begin{equation}
\mathcal{L}_{Q}
=
\frac{1}{B}
\sum_{b=1}^{B}
\left(\sum_i \hat{q}_{b,i}-Q_b\right)^2 ,
\end{equation}
where $Q_b=0$ if the total charge is unspecified.

To reflect the predominance of closed-shell states in condensed phases, we impose weak penalties on radical character and excessive charge:
\begin{align}
\mathcal{L}_{u}
&=
\frac{1}{|V_{\mathrm{valid}}|}
\sum_{i\in V_{\mathrm{valid}}}\hat{u}_i,\\
\mathcal{L}_{|q|}
&=
\frac{1}{|V_{\mathrm{valid}}|}
\sum_{i\in V_{\mathrm{valid}}}\left|\hat{q}_i\right|.
\end{align}
When ground-truth formal charges and unpaired electrons are available, their cross-entropy terms are also included:
\begin{align}
\mathcal{L}_{q\mathrm{CE}}
&=
\frac{1}{|V_{\mathrm{valid}}|}
\sum_{i\in V_{\mathrm{valid}}}
\mathrm{CE}(\mathbf{z}_i^{q},q_i^{\mathrm{gt}}),\\
\mathcal{L}_{u\mathrm{CE}}
&=
\frac{1}{|V_{\mathrm{valid}}|}
\sum_{i\in V_{\mathrm{valid}}}
\mathrm{CE}(\mathbf{z}_i^{u},u_i^{\mathrm{gt}}).
\end{align}
For formal-charge classification, the softmax is restricted to chemically allowed charge states for each element and unpaired-electron state, thereby preventing probability mass from being assigned to impossible states.
The weights are fixed throughout training at $(\mathrm{cons},Q,u,|q|,q\mathrm{CE},u\mathrm{CE})=(0.5,0.1,0.1,0.1,1.0,1.0)$.

\subsection{Data curation and datasets}

We used a classical-MD dataset for pretraining and a uMLIP trajectory dataset for fine-tuning and downstream evaluation. Both datasets comprised the same 128 nonreactive, topology-preserving condensed-phase systems under PBCs, enabling us to assess transfer across trajectory sources without changing the underlying chemical set. System sizes ranged from approximately 1000 to 3000 atoms, and both classical-MD and uMLIP trajectories were generated at 300 K and 1 atm. The classical-MD dataset contained approximately 1 million frames, and the uMLIP dataset contained approximately 25,000 frames.

For bookkeeping, the 128 systems were grouped into four mutually exclusive categories: 109 PURELIQUID, 5 SOLUTION, 9 ION, and 5 RADICAL systems. PURELIQUID comprised neat single-component condensed phases of neutral closed-shell molecules spanning diverse organic chemistries. SOLUTION comprised multicomponent solution-series systems, including water/ethanol mixtures and aqueous biomolecular solutes; end-member compositions in these series were retained in this category to maintain consistent dataset organization. ION comprised aqueous electrolyte solutions of NaCl, CaCl$_2$, and NaOH at 0.1, 1, and 10 M. RADICAL comprised systems containing open-shell species, namely allyl, benzyl, phenoxyl, TEMPO, and tert-butyl radicals. These labels were used only for dataset summary and analysis rather than as a formal chemical ontology. Complete system lists, exact dataset counts, and additional methodological details are provided in the Supporting Information.

\subsection{Reference annotation and data splits}

For each system $c$, the reference topology is defined as
\begin{equation}
\mathcal{T}_c^{\mathrm{ref}}=(\mathcal{V}_c,\mathcal{B}_c,\beta_c,q_c,u_c),
\end{equation}
where $\mathcal{B}_c$ is the set of undirected atom pairs that are covalently bonded, $\beta_c$ assigns their covalent states (Single, Double, Triple, Aromatic), and $q_c$ and $u_c$ denote atomwise formal charges and unpaired electrons. The reference topology for each system was extracted once from the RDKit molecule generated from the input SMILES and shared across all frames\cite{Landrum2016RDKit2016_09_4,kim2015universal}. Bond information was read from bond objects, whereas formal charges and unpaired-electron counts were read from atom attributes. The same procedure was used for both datasets.

For each snapshot $t$, the undirected set of proximity pairs is constructed as
\begin{equation}
\mathcal{P}_c^{(t)}=
\left\{\{i,j\}\,\middle|\, i<j,\ d_{ij}^{(t)}<r_c\right\}.
\end{equation}
Each pair $\{i,j\}\in\mathcal{P}_c^{(t)}$ is assigned the reference state
\begin{equation}
\ell_{ij}^{*(t)}=
\begin{cases}
\beta_c(i,j), & \{i,j\}\in\mathcal{B}_c,\\
\varnothing, & \{i,j\}\in\mathcal{P}_c^{(t)}\setminus\mathcal{B}_c.
\end{cases}
\end{equation}
Thus, a proximity pair receives a covalent state if it belongs to the reference topology; otherwise, it receives the noncovalent state $\varnothing$. Atomwise labels are defined as
\begin{equation}
q_i^{*(t)}=q_c(i),\qquad
u_i^{*(t)}=u_c(i),
\end{equation}
and are inherited across frames through the conserved atom indexing. During training, each undirected pair-state label is duplicated onto the two directed edges $(i,j)$ and $(j,i)$ handled by the GNN.

At evaluation time, the proximity-pair set is generated with the same cutoff, $r_c=3.0~\text{\AA}$. No special handling is introduced to add back reference bonds outside this cutoff.

For the classical MD dataset, training and validation were split at the frame level within each system (90\%/10\%). The same system therefore appeared in both sets, but no frame was duplicated. This design emphasizes robustness to thermal fluctuations and local environmental variation rather than extrapolation to unseen systems.

For the main uMLIP transfer evaluation reported in Table~\ref{tab:comparison_main}B, up to 200 frames were sampled from each system; 10\% were used for fine-tuning and the remaining 90\% for hold-out evaluation. The evaluation therefore probes robustness on unseen frames from the same systems and adaptation to the uMLIP distribution rather than extrapolation to unseen systems.

\subsection{Training protocol}

Training was performed in two stages to transfer a general condensed-phase representation from classical-MD data to uMLIP trajectories. We used AdamW with a batch size of 8 throughout\cite{zhou2024towards}. The model was first pretrained on the classical-MD dataset for 1 epoch with a learning rate of $2\times10^{-4}$. For the main transfer evaluation in Table~\ref{tab:comparison_main}B, the pretrained model was then fine-tuned on the uMLIP fine-tuning subset for 10 epochs using a learning rate of $1\times10^{-4}$, while all other hyperparameters were kept unchanged.

For the controlled ablation study in Table~\ref{tab:component_analysis}, we used a separate low-data adaptation protocol to compare model components under matched conditions. For each compound, 500 snapshots were sampled from the uMLIP trajectories; 90\% were used for fine-tuning and 10\% for evaluation, and each ablation variant was trained for 5 epochs from the same pretrained initialization. The values in Table~\ref{tab:component_analysis} should therefore be interpreted comparatively across rows and are not directly comparable to the full-set uMLIP results in Table~\ref{tab:comparison_main}B.

\subsection{Evaluation metrics and baselines}

We evaluated topology reconstruction from three perspectives: local classification accuracy, practical completion, and snapshot-level chemical validity. Although CoTAR predicts pair states in the set $\mathcal{C}=\{\varnothing\}\cup\mathcal{C}_{\mathrm{cov}}$, the main figures also report the derived metrics Bond and Bond Order for readability.

\paragraph{Local metrics.}
For each proximity pair, we define the binary reference label
\begin{equation}
a_{ij}^{*(t)}=\mathbf{1}\!\left[\ell_{ij}^{*(t)}\neq\varnothing\right],
\end{equation}
and let the predicted pair state and binary bond label be
\begin{equation}
\hat{\ell}_{ij}^{(t)}=\arg\max_{c\in\mathcal{C}}\hat p_{ij}^{(t)}(c),\qquad
\hat a_{ij}^{(t)}=\mathbf{1}\!\left[\hat{\ell}_{ij}^{(t)}\neq\varnothing\right].
\end{equation}
Bond is the F1 score for bonded versus nonbonded classification over undirected proximity pairs $\{i,j\}\in\mathcal{P}_c^{(t)}$. Bond Order is the weighted F1 score over the four covalent classes,
\[
\mathcal{C}_{\mathrm{cov}}=\{\mathrm{Single},\mathrm{Double},\mathrm{Triple},\mathrm{Aromatic}\},
\]
restricted to pairs for which both the reference and the prediction are covalent:
\begin{equation}
\mathcal{P}_{\mathrm{BO},c}^{(t)}=
\left\{\{i,j\}\in\mathcal{P}_c^{(t)}\ \middle|\ 
\ell_{ij}^{*(t)}\in\mathcal{C}_{\mathrm{cov}},\ 
\hat{\ell}_{ij}^{(t)}\in\mathcal{C}_{\mathrm{cov}}
\right\}.
\end{equation}
Accordingly, when a true covalent bond is predicted as $\varnothing$, it contributes a false negative to the Bond metric but is excluded from the Bond Order evaluation. Formal Charge (13 classes) and Unpaired Electron (2 classes) are also evaluated by weighted F1 scores.

\paragraph{Snapshot-level metrics.}
For high-throughput use, Success is the fraction of frames for which inference completes within the prescribed time limit and a topology is returned. To assess downstream usability, we introduce two per-snapshot metrics based on RDKit chemical-sanity diagnostics. Let $\hat{M}_i$ denote the molecular object reconstructed from the prediction, and let $n_{\mathrm{chem}}(\hat{M}_i)$ be the number of chemically problematic features reported by RDKit. If bond order, formal charge, or unpaired electrons remain undetermined because of failure or timeout, these unknown assignments are counted as downstream failures and added to the total problem count:
\begin{align}
n^{(i)}_{\mathrm{cp}}
&= n_{\mathrm{chem}}(\hat{M}_i)
 + \sum_{t\in\{\mathrm{BO},\mathrm{FC},\mathrm{UE}\}} n^{(i)}_{\mathrm{unk},t}.
\end{align}
We then define
\begin{align}
\mathrm{Valid}
&=
\frac{1}{N}\sum_{i=1}^{N}\mathbf{1}\!\left[n^{(i)}_{\mathrm{cp}}=0\right]
\end{align}
as the fraction of frames with neither RDKit-detected chemical problems nor unknown assignments, and
\begin{align}
\mathrm{Ave.\ Problems}
&=
\frac{1}{N}\sum_{i=1}^{N} n^{(i)}_{\mathrm{cp}}
\end{align}
as the average number of diagnostic problems and unknown assignments per snapshot. We emphasize Valid because downstream use requires a topology that is chemically complete and internally consistent, not merely locally accurate on pairwise labels. As the distance heuristic does not predict formal charges or unpaired electrons, these missing values are counted as unknown assignments when computing Valid.

\paragraph{Aggregation.}
For Table~\ref{tab:comparison_main}, the category-specific values of Bond, Bond Order, Formal Charge, and Unpaired Electron are computed by pooling confusion matrices over all systems within each category and recomputing the F1 scores from the pooled matrices. In contrast, Valid and Ave.\ problems are computed by concatenating per-snapshot values within each category and averaging them. The Total values in Table~\ref{tab:comparison_main} are computed over the full evaluation set and therefore are not arithmetic means of the category-specific values.

\paragraph{Baselines.}
We compared CoTAR with (i) a vdW-radius heuristic, with and without HMM smoothing, and (ii) a valence-rule-based baseline denoted RDKit.

For the vdW baseline, a pair is classified as covalently bonded when
\begin{equation}
\mathcal{P}_{\mathrm{vdW}}^{(t)}=
\left\{\{i,j\}\,\middle|\,
d_{ij}^{(t)}<
f\left(R_i^{\mathrm{vdW}}+R_j^{\mathrm{vdW}}\right)
\right\},
\end{equation}
where $R_i^{\mathrm{vdW}}$ denotes the element-specific van der Waals radius (in \AA) of atom $i$ taken from the Mendeleev database\cite{mendeleev2014}. For consistency with the vdW prior in Eq.~(19), the scaling parameter is fixed to $f=0.6$ in this baseline.
This baseline does not predict bond order, formal charge, or unpaired electrons; therefore, these quantities are treated as unknown when computing Valid. For vdW+HMM, HMM smoothing is applied to the same binary bonded/nonbonded state. All distance calculations under PBCs use the minimum-image convention.

For the RDKit baseline, the built-in \texttt{DetermineBonds} procedure is not directly applicable because it does not support PBCs. We therefore first construct a PBC-aware candidate-neighbor set from minimum-image distances and then assign bond orders by an xyz2mol-style valence-constrained procedure\cite{kim2015universal}. The candidate set is
\begin{equation}
\mathcal{P}_{\mathrm{cand}}^{(t)}=
\left\{\{i,j\}\,\middle|\,
d_{ij}^{(t)}<
\eta_{\mathrm{cov}}\left(r_i^{\mathrm{cov}}+r_j^{\mathrm{cov}}\right)
\right\},
\qquad
\eta_{\mathrm{cov}}=1.3,
\end{equation}
where $r_i^{\mathrm{cov}}$ denotes the element-specific covalent radius (in \AA) of atom $i$, taken from the Mendeleev database\cite{mendeleev2014}.
A per-snapshot time limit of $\tau=5$ s is imposed, and incomplete or failed assignments are treated as unknown.

\paragraph{Downstream validation.}
To test whether the predicted topologies are usable in downstream simulations, we connected the bond graph, bond orders, and formal charges predicted by fine-tuned CoTAR to the OpenFF~2.0.0 force field and performed short classical MD simulations in OpenMM\cite{boothroyd2023development,eastman2023openmm}. The main-text downstream results are reported for the full CoTAR pipeline after HMM smoothing. For ablation, we additionally compared the same predictions before and after HMM smoothing using frame~0 of each trajectory. This evaluation examines whether (i) the simulation can be initialized and propagated without divergence (MD feasibility) and (ii) the resulting density is consistent with that obtained from the reference topology. The same partial-charge assignment procedure was applied to both the reference and reconstructed topologies. Detailed MD settings and the exact dataset composition are provided in the Supporting Information.

\section{Results and Discussion}

We first examine computational cost, then evaluate predictive accuracy and the contribution of individual components. Finally, the usability of the reconstructed topologies in downstream classical MD is assessed. Unless otherwise noted, CoTAR denotes the full pipeline with the vdW prior, chemical constraints, and HMM smoothing enabled. In Table~\ref{tab:comparison_main}, CoTAR denotes the pretrained model for the classical-MD evaluation set and the fine-tuned model for the uMLIP evaluation set.

\subsection{Basic predictive accuracy}

Table~\ref{tab:comparison_main} summarizes the category-resolved and overall results on classical-MD and uMLIP evaluation sets. Bond connectivity was saturated or near-saturated across methods; hence, the more informative quantities are bond-order assignment and the snapshot-level metrics.

On the classical-MD evaluation set (Table~\ref{tab:comparison_main}A), CoTAR achieved an overall bond-order F1 of 0.906 and 94.0\% valid snapshots. The vdW-based heuristics completed on essentially all snapshots but, by construction, did not assign bond order, formal charge, or unpaired electrons; this limitation translated into low snapshot validity despite complete connectivity assignment. RDKit produced bond orders, but its overall bond-order F1 remained low (0.277), and incomplete or chemically problematic assignments led to only 21.0\% valid snapshots and 1832 average problems per snapshot. CoTAR was nearly perfect for PURELIQUID, SOLUTION, and ION systems, with category-wise valid-snapshot rates of 99.9\%, 100.0\%, and 99.4\%, respectively. RADICAL systems remained the most difficult category. In this category, vdW+HMM yielded a slightly higher valid-snapshot rate than CoTAR (83.3\% versus 76.5\%), but CoTAR still provided considerably better bond-order assignment (0.722 versus not applicable) and substantially fewer problems per snapshot (0.68 versus 3.0).

On the uMLIP evaluation set (Table~\ref{tab:comparison_main}B), CoTAR achieved an overall bond-order F1 of 0.911 and 84.7\% valid snapshots, compared with 36.6\% for the vdW-based heuristics and 19.7\% for RDKit. When the classical-MD-pretrained model was applied to uMLIP trajectories without adaptation, the valid-snapshot rate was 38.6\%; few-shot fine-tuning therefore more than doubled snapshot-level usability. Performance remained strongest for PURELIQUID and SOLUTION systems, with valid-snapshot rates of 95.1\% and 96.6\%, respectively, whereas ION and RADICAL systems remained more challenging at 67.4\% and 79.6\%. Even in these more difficult categories, the average number of problems per snapshot remained low for CoTAR (1.4 for ION and 1.8 for RADICAL) and far below those of the baselines. Taken together, these results show that the fine-tuned CoTAR model preserves strong local accuracy on uMLIP trajectories while substantially improving snapshot-level chemical usability relative to both heuristic and rule-based baselines.

\begin{table*}[t]
\centering
\caption{
Comparison of baseline methods and CoTAR on classical-MD and uMLIP evaluation sets.
CoTAR denotes the pretrained model in panel A and the fine-tuned model in panel B.
The uMLIP results in panel B correspond to the full transfer-evaluation protocol described in Methods.
Bond-connectivity F1 is omitted because it was saturated or near-saturated across methods.
}
\label{tab:comparison_main}
\small
\setlength{\tabcolsep}{3pt}
\renewcommand{\arraystretch}{1.03}
\begin{tabular}{@{}l@{\hspace{0.35em}}c@{\hspace{0.55em}}cccc@{}}
\toprule
& \multicolumn{1}{c}{\textbf{Total}} & \multicolumn{4}{c}{\textbf{By category}} \\
\cmidrule(lr){2-2}\cmidrule(l){3-6}
Method & Total & PURELIQUID & SOLUTION & ION & RADICAL \\
\midrule

\multicolumn{6}{@{}l}{\textbf{A. Classical-MD evaluation set}} \\
\multicolumn{6}{@{}l}{\emph{Bond-order F1} ($\uparrow$; vdW methods: N/A)} \\
RDKit & 0.277 & 0.409 & 0.262 & 0.000 & 0.436 \\
CoTAR & \textbf{0.906} & \textbf{0.903} & \textbf{1.000} & \textbf{1.000} & \textbf{0.722} \\

\addlinespace[1pt]
\multicolumn{6}{@{}l}{\emph{Valid snapshots (\%)} ($\uparrow$)} \\
vdW         & 45.9 & 89.0 & 11.7 & 0.0 & 83.0 \\
vdW + HMM   & 52.0 & 93.1 & 26.6 & 4.9 & \textbf{83.3} \\
RDKit       & 21.0 & 30.5 & 20.0 & 0.0 & 33.3 \\
CoTAR       & \textbf{94.0} & \textbf{99.9} & \textbf{100.0} & \textbf{99.4} & 76.5 \\

\addlinespace[1pt]
\multicolumn{6}{@{}l}{\emph{Avg. problems/snapshot} ($\downarrow$)} \\
vdW         & 13 & 1.7 & 20 & 28 & 3.0 \\
vdW + HMM   & 6 & 0.95 & 6.9 & 13 & 3.0 \\
RDKit       & 1{,}832 & 1{,}982 & 2{,}355 & 2{,}070 & 923 \\
CoTAR       & \textbf{0.17} & \textbf{$<10^{-3}$} & \textbf{0} & \textbf{0.006} & \textbf{0.68} \\

\midrule

\multicolumn{6}{@{}l}{\textbf{B. uMLIP evaluation set}} \\
\multicolumn{6}{@{}l}{\emph{Bond-order F1} ($\uparrow$; vdW methods: N/A)} \\
RDKit & 0.266 & 0.290 & 0.573 & 0.000 & 0.202 \\
CoTAR & \textbf{0.911} & \textbf{0.902} & \textbf{1.000} & \textbf{1.000} & \textbf{0.742} \\

\addlinespace[1pt]
\multicolumn{6}{@{}l}{\emph{Valid snapshots (\%)} ($\uparrow$)} \\
vdW         & 36.6 & 70.2 & 0.0 & 0.0 & 76.0 \\
vdW + HMM   & 36.6 & 70.2 & 0.0 & 0.0 & 76.0 \\
RDKit       & 19.7 & 18.8 & 40.0 & 0.0 & 20.0 \\
CoTAR       & \textbf{84.7} & \textbf{95.1} & \textbf{96.6} & \textbf{67.4} & \textbf{79.6} \\

\addlinespace[1pt]
\multicolumn{6}{@{}l}{\emph{Avg. problems/snapshot} ($\downarrow$)} \\
vdW         & 90 & 8.0 & 135 & 209 & 6.4 \\
vdW + HMM   & 90 & 8.0 & 135 & 209 & 6.4 \\
RDKit       & 2{,}044 & 2{,}279 & 1{,}343 & 2{,}246 & 2{,}309 \\
CoTAR       & \textbf{0.86} & \textbf{0.19} & \textbf{0.034} & \textbf{1.4} & \textbf{1.8} \\

\bottomrule
\end{tabular}
\end{table*}

\subsection{Ablation study}

Table~\ref{tab:component_analysis} summarizes the ablation study in a controlled low-data adaptation setting. For each compound, 500 uMLIP snapshots were sampled, 90\% were used for fine-tuning and 10\% for evaluation, and each variant was trained for 5 epochs. As this protocol differs from the full uMLIP transfer evaluation in Table~\ref{tab:comparison_main}B, Table~\ref{tab:component_analysis} is intended to isolate relative component contributions rather than reproduce the absolute full-set numbers.

The effect of HMM smoothing can be assessed directly by comparing the first two rows. Disabling HMM reduced the valid-snapshot rate only modestly, from 98.9\% to 97.7\%, and increased the average number of problems per snapshot from 0.02 to 0.04. In this setting, HMM acts primarily as a final cleanup step rather than as the main source of chemical consistency.

The stronger effect came from the chemical constraints. With HMM disabled, retaining the chemical constraints but removing the vdW prior still yielded 97.4\% valid snapshots and 0.24 average problems per snapshot. By contrast, configurations without chemical constraints showed markedly worse snapshot usability, with 95.0--95.3\% valid snapshots and 3.07--6.23 average problems per snapshot. The formal-charge F1 scores followed the same pattern, decreasing from 0.9972 with chemical constraints active to 0.9899 without them.

The vdW prior had a smaller and more context-dependent effect. When chemical constraints were active and HMM was disabled, adding the vdW prior slightly improved the valid-snapshot rate (97.4\% to 97.7\%) and reduced the average number of problems per snapshot (0.24 to 0.04), while leaving the local F1 scores similar. By contrast, without chemical constraints, the vdW prior did not improve the outputs and increased the average number of problems per snapshot from 3.07 to 6.23. Overall, Table~\ref{tab:component_analysis} indicates that the use of chemical constraints are the main source of chemically usable assignments. HMM provides an additional refinement, and the vdW prior is most useful when combined with the constraints rather than used as a standalone correction.

\begin{table*}[t]
\centering
\caption{
Ablation study of CoTAR in a controlled low-data adaptation setting on uMLIP trajectories.
For each compound, 500 snapshots were sampled; 90\% were used for fine-tuning and 10\% for evaluation, and each variant was fine-tuned for 5 epochs from the same pretrained model.
Enabled components are indicated by checkmarks, and blank cells indicate disabled components.
vdW = van der Waals prior; Chem. = chemical constraints.
BO, FC, and UE denote weighted F1 scores for bond order, formal charge, and unpaired electron, respectively.
Bond connectivity is omitted because it was saturated at 1.000 in all settings.
The absolute values are intended for relative comparison among ablation variants and are not directly comparable to the full-set uMLIP results in Table~\ref{tab:comparison_main}B.
}
\label{tab:component_analysis}
\small
\setlength{\tabcolsep}{4pt}
\renewcommand{\arraystretch}{1.02}
\begin{tabular}{@{}ccc@{\hspace{0.8em}}rrrrr@{}}
\toprule
\multicolumn{3}{c}{Comp.} & \multicolumn{5}{c}{Metrics} \\
\cmidrule(lr){1-3}\cmidrule(l){4-8}
vdW & Chem. & HMM & BO $\uparrow$ & FC $\uparrow$ & UE $\uparrow$ & Valid (\%) $\uparrow$ & Avg. prob. $\downarrow$ \\
\midrule
\checkmark & \checkmark & \checkmark & \textbf{0.9061} & \textbf{0.9966} & \textbf{0.9999} & \textbf{98.9} & \textbf{0.02} \\
\checkmark & \checkmark &             & 0.9015          & 0.9911          & 0.9999          & 97.7          & 0.04          \\
            & \checkmark &             & 0.9061          & 0.9972          & 1.0000          & 97.4          & 0.24          \\
\checkmark &             &             & 0.9063          & 0.9899          & 0.9998          & 95.0          & 6.23          \\
            &             &             & 0.9062          & 0.9899          & 0.9998          & 95.3          & 3.07          \\
\bottomrule
\end{tabular}
\end{table*}

\subsection{Reconnection to classical MD}

\begin{figure}[t]
  \centering
  \maybeincludegraphics[width=0.9\columnwidth]{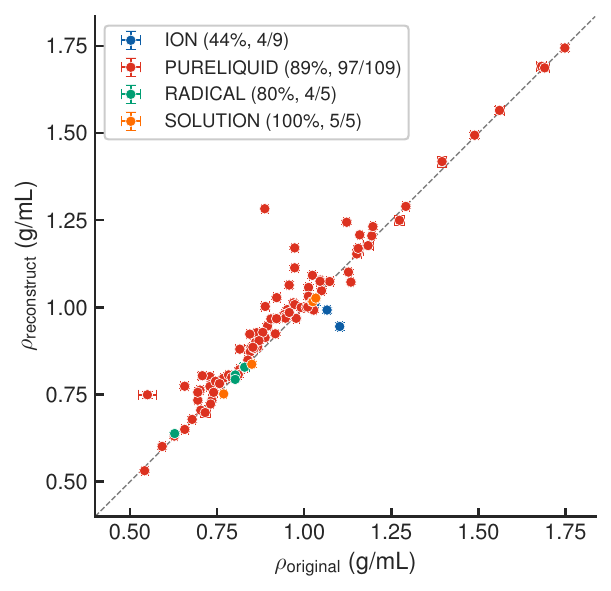}
  \caption{Downstream classical-MD validation using topologies predicted by the full CoTAR pipeline. The horizontal axis is the density obtained from the reference topology, $\rho_{\mathrm{original}}$, and the vertical axis is the density obtained from the predicted topology, $\rho_{\mathrm{reconstruct}}$. Points are colored by category, error bars denote $\pm 1\sigma$, and the legends report category-specific MD-feasibility rates.}
  \label{fig:density_comparison}
\end{figure}

We finally tested whether topologies predicted by the full CoTAR pipeline could be reconnected to a classical force field and propagated by short classical-MD runs. Across the 128 uMLIP systems, 110 systems (85.9\%) completed the downstream validation (Figure~\ref{fig:density_comparison}). The success rate was highest for SOLUTION systems (5/5) and remained high for PURELIQUID (97/109) and RADICAL (4/5) systems, whereas ION systems were the most difficult (4/9). HMM smoothing modestly improved system-level MD simulation feasibility from 107/128 (83.6\%) to 110/128 (85.9\%), consistent with its role as a temporal cleanup step rather than the primary source of chemical consistency.

For systems that completed the validation, the densities obtained from CoTAR-reconstructed topologies closely tracked those obtained from the reference topologies over a broad density range and the points clustered near the identity line with no obvious systematic bias (Figure~\ref{fig:density_comparison}). This agreement indicates that the reconstructed bond graphs, bond orders, and formal charges were generally sufficiently accurate for direct force-field assignment and short downstream simulation. The remaining failures were concentrated in ION systems, consistent with their lower snapshot-level accuracy (Table~\ref{tab:comparison_main}) and the stronger sensitivity of ionic systems to small residual errors in charge or connectivity.

\subsection{Inference speed}

\begin{figure}[t]
  \centering
  \maybeincludegraphics[width=\columnwidth]{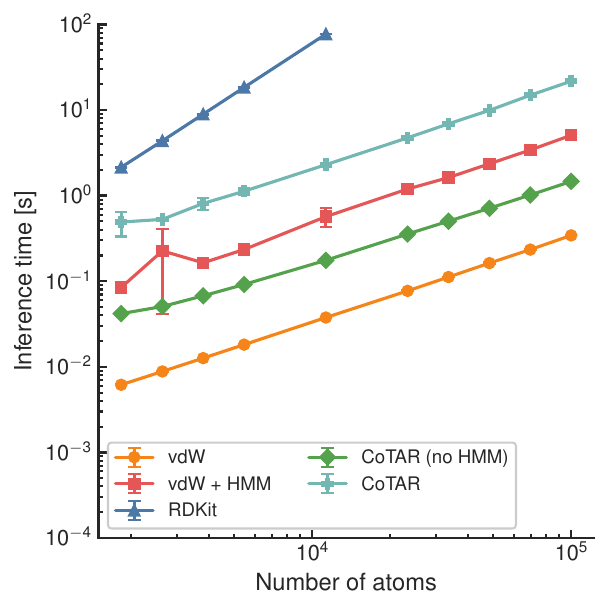}
  \caption{Inference time per snapshot versus atom count for a separate liquid-toluene scaling test. The atom count was varied by changing the number of toluene molecules in the periodic liquid cell. The same configurations were evaluated with all methods shown in the legend. This figure is intended only to illustrate computational scaling and is not part of the 128-system accuracy benchmark.}
  \label{fig:inference_time}
\end{figure}

Practical use in condensed phases requires stable execution at an acceptable computational cost. Figure~\ref{fig:inference_time} shows the inference time per snapshot as a function of atom count. The simple vdW heuristic was the fastest method over the full system-size range, as expected. CoTAR showed a smooth increase in cost with system size and remained practical for the condensed-phase systems examined. RDKit exhibited the steepest growth with atom count and became markedly more expensive on larger graphs. In dense periodic systems, this behavior is consistent with the increasing cost of rule-based bond-order assignment on large candidate graphs.

\subsection{Interpretation, scope, and limitations}

The present benchmark should be interpreted within a defined domain of applicability. First, the study is restricted to nonreactive, topology-preserving condensed-phase systems. Second, the classical-MD benchmark uses a frame-level split within each system, and the main uMLIP evaluation uses same-system few-shot fine-tuning followed by hold-out evaluation on the remaining frames. The reported gains therefore establish robustness to thermal fluctuations, local environmental variation, and simulation-modality shift within the represented systems, but they do not yet establish zero-shot extrapolation to entirely unseen chemistries or topologies.

The large improvement on uMLIP trajectories after few-shot fine-tuning (38.6\% to 84.7\% valid snapshots) is nevertheless practically important. It indicates that CoTAR is effective when a modest amount of adaptation data can be obtained for the target system, which is a realistic use case for production uMLIP workflows. At the same time, the current benchmark is dominated by PURELIQUID systems (109 of 128 systems); the ION and RADICAL categories each contain fewer than ten systems. Hence, category-level trends in these minor classes should be interpreted cautiously.

The remaining failures in ION and RADICAL systems are chemically plausible. Ionic systems are more sensitive to small residual errors in charge localization and intermolecular-contact classification because such errors can propagate directly into force-field assignment, while open-shell systems impose additional consistency requirements on bond orders, formal charges, and unpaired-electron counts. Finally, the downstream classical-MD validation should be interpreted as evidence that the reconstructed topologies are reconnectable and usable in short conventional MD runs; it does not by itself constitute a full validation of all force-field-dependent thermodynamic or kinetic properties.

\section{Conclusions}

In this study, we developed CoTAR, a hybrid GNN--HMM framework for reconstructing molecular topology and atom states from atomic coordinates in condensed phases. Across 128 nonreactive, topology-preserving systems, CoTAR outperformed both distance-based and rule-based baselines on classical-MD and uMLIP evaluation sets and achieved strong bond-order accuracy together with high snapshot-level validity.

The ablation study clarified the roles of the main components. Chemical constraints were the primary contributor to chemically usable assignments. HMM smoothing provided an additional improvement in consistency, and the vdW prior was most effective when used together with the constraints. The large gain after uMLIP few-shot fine-tuning further indicates that CoTAR is particularly useful when a modest amount of same-system adaptation data can be obtained. The reconstructed topologies also supported direct reconnection to classical MD for most systems and reproduced the density trends of the reference topologies in successful runs.

The present domain of applicability is nonreactive, topology-preserving condensed-phase trajectories and the chemistries represented in the current benchmark. The current benchmark does not yet establish zero-shot extrapolation to entirely unseen chemistries. Moreover, the downstream MD tests assess reconnectability and short-time simulation usability rather than exhaustive force-field validation. The most important next steps are improved robustness for ionic and open-shell systems, stronger evaluation on unseen chemical systems, and extension of the framework to reactive trajectories.

\section*{Data and Software Availability}

The full raw trajectories and the complete internal training/evaluation pipeline used in this study are proprietary and are not publicly released. To support verification and reuse of the reported methodology, a public GitHub repository will be made available at \url{https://github.com/pfnet-research/cotar}. The repository will contain a minimal validation dataset, training scripts, trained model parameters, inference code, configuration files, and documentation sufficient to reproduce a representative validation workflow and run CoTAR on user-provided coordinate trajectories. Additional nonconfidential materials may be available from the corresponding author upon reasonable request, subject to permission from the data owner where applicable.

\clearpage
\begin{suppinfo}
Supporting Information: complete system list for the 128 systems used in this study; system-level dataset composition; benchmark split design and 
interpretation; controlled low-data protocol for the Table~\ref{tab:component_analysis} ablation study; and downstream classical-MD validation settings (PDF).
\end{suppinfo}

\begin{acknowledgement}
We thank So Takamoto (Preferred Networks, Inc.), Ryosuke Kuwabara (MITSUI \& CO., LTD.), Sannohe Kunio (IBLC Co., Ltd.), and their colleagues for helpful discussions and comments.
\end{acknowledgement}

\bibliography{references}
\end{document}


\renewcommand{\thetable}{S\arabic{table}}
\renewcommand{\thefigure}{S\arabic{figure}}

\section{Complete system list and dataset composition}

Table~\ref{tab:system_list} lists the 128 systems used in this study. To improve scanability, the identifiers are grouped by category and arranged in two columns while preserving the exact internal dataset naming convention used during data generation and evaluation.

\footnotesize
\setlength{\tabcolsep}{4pt}
\renewcommand{\arraystretch}{1.05}
\begin{longtable}{@{}>{\raggedright\arraybackslash}p{0.49\textwidth}>{\raggedright\arraybackslash}p{0.49\textwidth}@{}}
\caption{Complete list of the 128 systems used in this study, grouped by category.}\label{tab:system_list}\\
\toprule
\midrule
\endfirsthead
\multicolumn{2}{@{}l}{\tablename\ \thetable\ (continued)}\\
\toprule
\midrule
\endhead
\midrule
\multicolumn{2}{r}{Continued on next page}\\
\endfoot
\bottomrule
\endlastfoot
\addlinespace[2pt]
\multicolumn{2}{@{}l}{\textbf{PURELIQUID} (109 systems)} \\
\cmidrule(l){1-2}
\path{1_1_1_trichloro_2_2_2_trifluoroethane} & \path{1_1_1_trichloroethane} \\
\path{1_1_2_trichloro_1_2_2_trifluoro_ethane} & \path{1_1_2_trichloroethylene} \\
\path{1_1_dichloroethane} & \path{1_1_dichloroethene} \\
\path{1_2_dichlorobenzene} & \path{1_2_dichloroethane} \\
\path{1_3_dioxolan_2_one} & \path{1_4_diethylbenzene} \\
\path{1_4_dioxane} & \path{1_methyl_2_pyrrolidone} \\
\path{1_nitroethane} & \path{2_2_2_hydroxyethoxyethoxyethanol} \\
\path{2_2_4_trimethylpentane} & \path{2_2_butoxyethoxyethanol} \\
\path{2_2_ethoxyethoxyethanol} & \path{2_2_hydroxyethoxyethanol} \\
\path{2_2_methoxyethoxyethanol} & \path{2_2_methoxypropoxypropan_1_ol} \\
\path{2_6_dimethylheptan_4_one} & \path{2_aminoethanol} \\
\path{2_butoxyethanol} & \path{2_ethoxyethanol} \\
\path{2_hydroxybenzoicacidmethylester} & \path{2_methylpropan_1_ol} \\
\path{2_methylpropionicacidisobutylester} & \path{2_nitropropane} \\
\path{3_5_5_trimethylcyclohex_2_en_1_one} & \path{4_hydroxy_4_methyl_pentan_2_one} \\
\path{4_methyl_1_3_dioxol_2_one} & \path{4_methylpentan_2_one} \\
\path{5_methylhexan_2_one} & \path{aceticacid} \\
\path{aceticacid2_ethoxyethylester} & \path{aceticacidbutylester} \\
\path{aceticacidethylester} & \path{aceticacidisoamylester} \\
\path{aceticacidisobutylester} & \path{aceticacidmethylester} \\
\path{acetone} & \path{acetophenone} \\
\path{aminopropyltrimethoxysilane} & \path{aniline} \\
\path{benzene} & \path{benzoicacid} \\
\path{butan_1_ol} & \path{butan_2_ol} \\
\path{butan_2_one} & \path{carbondisulfide} \\
\path{carbonicaciddiethylester} & \path{carbontetrachloride} \\
\path{chlorobenzene} & \path{chloroform} \\
\path{chloromethane} & \path{cyclohexane} \\
\path{cyclohexanol} & \path{cyclohexanone} \\
\path{cyclohexylamine} & \path{decan_1_ol} \\
\path{dichloromethane} & \path{dodecane} \\
\path{ethane_1_2_diamine} & \path{ethanethiol} \\
\path{ethanol} & \path{ethoxyethane} \\
\path{ethylbenzene} & \path{ethyleneglycol} \\
\path{formamide} & \path{formicacid} \\
\path{formicacidethylester} & \path{glycerol} \\
\path{m_cresol} & \path{methanol} \\
\path{methylcyclohexane} & \path{methylsulfinylmethane} \\
\path{morpholine} & \path{n_butane} \\
\path{n_heptane} & \path{n_hexane} \\
\path{n_n_dimethylformamide} & \path{n_octane} \\
\path{n_pentane} & \path{naphthalene} \\
\path{nitrobenzene} & \path{nitromethane} \\
\path{o_xylene} & \path{octamethylcyclotetrasiloxane} \\
\path{oleicacid} & \path{pentan_3_one} \\
\path{pentyl_acetate} & \path{phenol} \\
\path{phenylmethanol} & \path{phenylmethoxymethylbenzene} \\
\path{phenyltriethoxysilane} & \path{prop_2_en_1_ol} \\
\path{propan_1_ol} & \path{propan_2_ol} \\
\path{propane_1_2_diol} & \path{pyridine} \\
\path{quinoline} & \path{resorcinol} \\
\path{stearicacid} & \path{styrene} \\
\path{sulfuricaciddiethylester} & \path{tetraethoxysilane} \\
\path{tetrahydrofuran} & \path{toluene} \\
\path{water} &  \\
\addlinespace[2pt]
\multicolumn{2}{@{}l}{\textbf{SOLUTION} (5 systems)} \\
\cmidrule(l){1-2}
\path{water_alanine_dipeptide} & \path{water_ethanol_0_100} \\
\path{water_ethanol_100_0} & \path{water_ethanol_50_50} \\
\path{water_glycine} &  \\
\addlinespace[2pt]
\multicolumn{2}{@{}l}{\textbf{ION} (9 systems)} \\
\cmidrule(l){1-2}
\path{cacl2_0.1M} & \path{cacl2_1M} \\
\path{cacl2_10M} & \path{nacl_0.1M} \\
\path{nacl_1M} & \path{nacl_10M} \\
\path{naoh_0.1M} & \path{naoh_1M} \\
\path{naoh_10M} &  \\
\addlinespace[2pt]
\multicolumn{2}{@{}l}{\textbf{RADICAL} (5 systems)} \\
\cmidrule(l){1-2}
\path{allyl_radical} & \path{benzyl_radical} \\
\path{phenoxyl_radical} & \path{tempo} \\
\path{tert_butyl_radical} &  \\
\end{longtable}
\normalsize
\setlength{\tabcolsep}{6pt}
\renewcommand{\arraystretch}{1.0}

\begin{table}[h]
  \caption{Exact dataset composition at the system level. The classical-MD dataset contained 980{,}563 frames, and the uMLIP dataset comprised the same 128 systems and 24{,}924 frames in total.}
  \label{tab:dataset_counts}
  \centering
  \begin{tabular}{lrrr}
    \toprule
    Category & Systems & Classical-MD frames & uMLIP systems \\
    \midrule
    PURELIQUID & 109 & 894{,}127 & 109 \\
    SOLUTION   & 5   & 41{,}015  & 5   \\
    ION        & 9   & 33{,}214  & 9   \\
    RADICAL    & 5   & 12{,}207  & 5   \\
    \midrule
    Total      & 128 & 980{,}563 & 128 \\
    \bottomrule
  \end{tabular}
\end{table}

The main-text uMLIP analyses use sampled snapshot subsets from these trajectories. The benchmark is intentionally diverse but imbalanced at the system level, with PURELIQUID systems comprising 109 of the 128 systems. Accordingly, category-level trends for the smaller ION and RADICAL subsets should be interpreted more cautiously than those for PURELIQUID systems.

\section{Benchmark split design and interpretation}

For the classical-MD benchmark, the split was performed at the frame level within each system (90\%/10\% train/validation). The same chemical systems therefore appear in both subsets, but no frame is duplicated. This benchmark is intended to test robustness to thermal fluctuations and local environmental variation rather than zero-shot extrapolation to unseen systems.

For the main uMLIP transfer evaluation reported in Table~1B of the main text, up to 200 snapshots were sampled per system. Of these, 10\% were used for few-shot fine-tuning and the remaining 90\% were used for hold-out evaluation. The resulting benchmark therefore measures adaptation to the uMLIP distribution and performance on unseen frames from the same systems, rather than transfer to entirely unseen chemistries.

\section{Controlled low-data protocol for the ablation study}

The ablation results in Table~2 of the main text were obtained with a separate controlled low-data adaptation protocol. For each compound, 500 snapshots were sampled from the uMLIP trajectories. Of these, 90\% were used for fine-tuning and 10\% for evaluation. All ablation variants were initialized from the same classical-MD-pretrained weights and fine-tuned for 5 epochs. Because this protocol differs from the full uMLIP transfer evaluation reported in Table~1B of the main text, the absolute values in Table~2 should be interpreted comparatively across model variants rather than compared directly with Table~1B.

\section{Simulation details for downstream classical-MD validation}
\label{sec:sim_detail}

Unless otherwise noted, the classical-MD simulations used for downstream validation were performed with OpenMM~8.1.1. This section summarizes the system construction, force-field assignment, and equilibration protocol used to connect the reference (GT) and CoTAR-predicted topologies to a classical force field and to test whether stable MD could be performed.

\subsection{Classical-MD setup}
\label{sec:classicmd}

In the downstream validation reported in the main text, the GT and CoTAR-predicted topologies were connected to a classical force field, and their stability under conventional classical MD was examined. Before these calculations, all systems were equilibrated with a classical force field to obtain physically reasonable initial configurations.

The OpenFF~2.0.0 force field was used for bonded and nonbonded interactions. For chemical environments not directly covered by the standard parameter set, auxiliary bonded parameters were generated by fitting to potential-energy surfaces obtained from PFP calculations\cite{Matlantis}. Partial charges were assigned by Bader charge analysis with PFP (v6.0.0) and uniformly scaled by a factor of 0.6. This procedure yielded charge values close to AM1-BCC charges and was adequate for the initial equilibration.

Electrostatic interactions were treated with the particle-mesh Ewald method. A real-space cutoff of 0.9~nm was used, with a switching function applied from 0.8~nm. van der Waals interactions were also truncated at 0.9~nm, and long-range dispersion corrections were applied to the energy and pressure.

The simulations were carried out in three stages: NVT equilibration (20{,}000 steps, 10~ps), NPT equilibration (300{,}000 steps, 150~ps), and NPT production (500{,}000 steps, 250~ps). The temperature was set to 300~K throughout, and the pressure was set to 1~atm during the NPT stages. Temperature control used a Langevin thermostat with a coupling constant of 1.0~ps$^{-1}$, and pressure control used a Monte Carlo barostat. A time step of 0.5~fs was used, and all bonds involving hydrogen atoms were constrained with the RATTLE algorithm.

The equilibrated structures were used as starting configurations for the short classical-MD runs employed in the downstream validation. The same protocol was applied to both the GT and CoTAR-predicted topologies.

\bibliography{references}